\documentclass[aps, prb, twocolumn, uperscriptaddress]{revtex4-1}
\usepackage{amsmath}
\usepackage{amsfonts,amssymb}
\usepackage{graphicx}
\usepackage{epstopdf}
\usepackage{hyperref}
\usepackage{color}
\usepackage{times}
\begin{document}	
\title{Non-Hermitian Chiral Edge Modes With Complex Fermi Velocity}
\author{Fei Yang}
\affiliation{Center for Advanced Quantum Studies, Department of Physics, Beijing Normal University, Beijing 100875, China}
\author{Xue-Ping Ren}
\affiliation{Center for Advanced Quantum Studies, Department of Physics, Beijing Normal University, Beijing 100875, China}
\author{Su-Peng Kou}
\email{spkou@bnu.edu.cn}
\affiliation{Center for Advanced Quantum Studies, Department of Physics, Beijing Normal University, Beijing 100875, China}

\begin{abstract}
Recently, much attention has been paid to uncovering the influence of dissipation on a quantum system, particularly on how the non-Hermitian (NH) terms modify the band topology of topological materials and reshape the profile of the wavefunctions of a system (or the NH skin effect).
In this paper, a specific NH skin effect that induced by local dissipation is studied for chiral edge modes, in which the NH term corresponds to the local imaginary Fermi velocity of the chiral edge modes.
By solving the NH Schr\"{o}dinger equation of the non-Hermitian chiral edge modes (nhCEs) with complex Fermi velocity, we uncovered the remarkable complex spectra and the wavefunctions of the nhCEs.
We find that the complex spectra of these modes is a straight line in the topological materials, and its chirality can separates the modes with positive energy from those with negative energy, in which they are localized at different positions.
We also studied the nhCEs at the boundary of 2-dimensional (2D) topological materials, the 2D $p$-wave superconductor and the Qi-Wu-Zhang model, in which the general law of nhCEs was verified.
We expect that our findings will pave the way for researching the transport properties of the chiral edge modes in the non-equilibrium context.
\end{abstract}
\pacs{11.30. Er, 75.10. Jm, 64.70. Tg, 03.65.-W}
\maketitle

\section{Introduction}
The physics of non-Hermitian (NH) systems have attracted much attention in recent years \cite{NH_system1,NH_system2}, not only because of its remarkable topological property \cite{NH_topology1,NH_topology2}, but also because of many interesting phenomena that do not have Hermite counterpart, such as $\mathcal{PT}$-physics \cite{NH_PT1,NH_PT2,NH_PT3} and NH skin effect\cite{NH_skin0,NH_skin1,NH_skin2,NH_skin3,NH_skin4}.
The NH Hamiltonian is an effective description of some open quantum systems \cite{open_system_NH} or the system that composed of quansi-particles \cite{finite_lifetime}.
Many remarkable phenomena has been uncovered by studying the effective NH Hamiltonian, such as unidirectional invisibility \cite{uni_directional1,uni_directional2}, single-mode lasing \cite{sigle_mode_pt1,single_mode_pt2}, and enhanced sensitivity \cite{enhenced_apply1_pt,enhenced_apply2_pt,enhenced_apply3_pt,enhenced_apply4_pt}.
In a NH system, the main problem is to figure out how the dissipation (or the NH terms) affects the property of the system, which are determined both by its complex spectra and its wavefunctions \cite{NH_spectra_vector,NH_BBC4}.
Fortunately, most of problems have been settled thanks to the non-Bloch band theory \cite{NH_skin3,NH_skin4,non_bloch}, which only takes into account the global dissipation in the bulk.
However, dissipation that only occurs in certain part of system is more common and easier to control in real experiments \cite{nonequilibrium_boundary,open_system_local}, which makes the problem that how the local dissipation affects the property of a system becomes an urgent issue needs to be addressed.
Many interesting phenomena induced by local dissipation have been revealed \cite{local_NH1,local_NH2,local_NH3}, such as scale-free localization \cite{scale_free_local,imputiry_nonreciprocal} and the $\mathcal{PT}$-physics \cite{local_nh_PT}.
The influence of local dissipation on the anomalous chiral edge modes is also an interesting topic, not only because of its amazing transport property, but also because it would gives insights to the problem of hybrid skin-topological effect \cite{chiral_localization,high_order_skin1}.

The anomalous chiral edge modes can only be found at the edge of topological materials because of the Nielsen-Ninomiya theorem \cite{NN_theorem1,NN_theorem2}.
In solid state materials, these anomalous edge modes serve as perfectly robust transport channels that produce quantised Hall conductance.
Due to the topological protection of the bulk band structure, they are robust against localization, meaning that it remains extended even though there is defect or disorder \cite{edge_localization1,edge_localization2,edge_localization3,edge_localization4}.
Incredibly, the inclusion of dissipation would make it possible to localize these chiral edge modes, which is related to the hybrid skin-topological effect \cite{chiral_localization}.
However, we find that previous studies of the non-Hermitian chiral edge modes (nhCEs) experiencing imaginary potential \cite{chiral_localization} can not be generalized to the chiral Majorana edge modes in the two-dimensional (2D) topological supercondutors (SC), that the chiral Majorana edge modes remained as the extended states due to the intrinsic particle-hole symmetry.
Fortunately, this problem is solved by considering the nhCEs with complex Fermi velocity, that the non-hermitian Majorana chiral edge modes become the localized states.

In this paper, the general properties of nhCEs with complex Fermi velocity is revealed by solving the NH Schr\"{o}dinger equation of nhCEs.
We find that the interplay between dissipation and the underlying low energy physics gives rise to non-trivial energy spectra and distinct localization rules, in which the chirality of edge modes is essential in determining the position where they are localized.
We find that the complex spectra of nhCEs with local complex Fermi velocity is a straight line, and the chirality can tells the difference between the modes with positive energy and the those with negative energy, in which they are localized at different positions.
Moreover, general properties of the nhCEs are verified in the 2-dimensional (2D) topological materials , the 2D $p$-wave SC and Qi-Wu-Zhang (QWZ) model.

The structure of this paper is as follows.
In Sec. II, we formulate the Shr\"{o}dinger equation of the nhCEs with local complex Fermi velocity, in which the influences of the dissipation to its wavefunction and its energies are revealed.
We find that its localization rules are related to the chirality of the edge modes, so called the chiral localization.
In Sec. III, we study the nhCEs at the boundary of 2D $p$-wave SC. It is find that the only NH term that would results in localized chiral Majorana edge modes is the imaginary Fermi velocity.
In Sec. IV, the nhCEs in the 2D QWZ model are studied.
Main findings and future directions are summarized in Sec. V.

\section{the non-Hermitian anomalous chiral edge modes}
The anomalous chiral edge modes can only exist at the boundary of topological materials due to the Nielsen-Ninomiya theorem \cite{NN_theorem1,NN_theorem2}, these long-wave-length excitations are restricted to propagate only in one direction and are protected by the non-trivial topology of bulk.
And the dispersion relation of those modes are linear in $k$, i. e. $\epsilon_k =  v_{f} k$, they are correspond to the robust edge channels in the Quantum Hall Liquids, that
\begin{equation}
	H_{\text{edge}} = v_{f} \hat{k},
\end{equation}
where $v_{f}$ is the Fermi velocity that defined as $\left.\frac{\partial\epsilon_k}{\partial k}\right|_{k = k_{f} }$.
And in the topological materials, the Fermi energy $\epsilon_{f}$ and Fermi wave-vector $k_{f}$ is set to zero for simplicity.
Furthermore, we need to set the upper bound of wave-vector $k_{t}$ to the anomalous edge modes in the actual topological system, in which the modes that wave-vector above $k_{t}$ is no longer the anomalous edge modes, in other words, the anomalous edge modes are those satisfy $|E_{\text{edge}}| < |v_{f} {k_{t}}|$.

Now, we want to know what the consequences would be if these topological modes are dissipative.
The physical meaning of dissipation is that the lifetime of edge excitations is finite, which is usually corresponds to the situations that either there is particle gain/loss at the boundary of system, or the boundary potential has a non-zero imaginary part.
Given that the bulk topology of system is unaffected by the local dissipation which only exist at the boundaries, then it is expected that only the anomalous chiral edge modes would experience dissipation,  and become the nhCES we want to study.
For the nhCEs, it is natural that the eigenenergys might have non-zero imaginary parts, which the effective low-energy description is a non-Hermitian Hamiltonian
\begin{equation}\label{Eff_edge}
	H_{\text{nh},\text{edge}} = H_{\text{edge}} + i\hat{\gamma},
\end{equation}
where the eigenenergy's imaginary part is depend on how $i\hat{\gamma}$ is varied.
Furthermore, because the low energy physics is linear in $k$, then there would have two kinds of dissipation, one is $k$-dependent (that $\hat{\gamma}$ is linear in $\hat{k}$), another one is $k$-independent.
The $k$-independent dissipation corresponds to the gain/loss of particles and has been studied in Ref. \cite{chiral_localization}, we will study the $k$-dependent dissipation in this paper, which the dissipation is the non-reciprocal hopping at the boundary of system and corresponds to the imaginary fermi velocity of the anomalous chiral edge modes.

\begin{figure}[h]
	\includegraphics[width=0.5\textwidth]{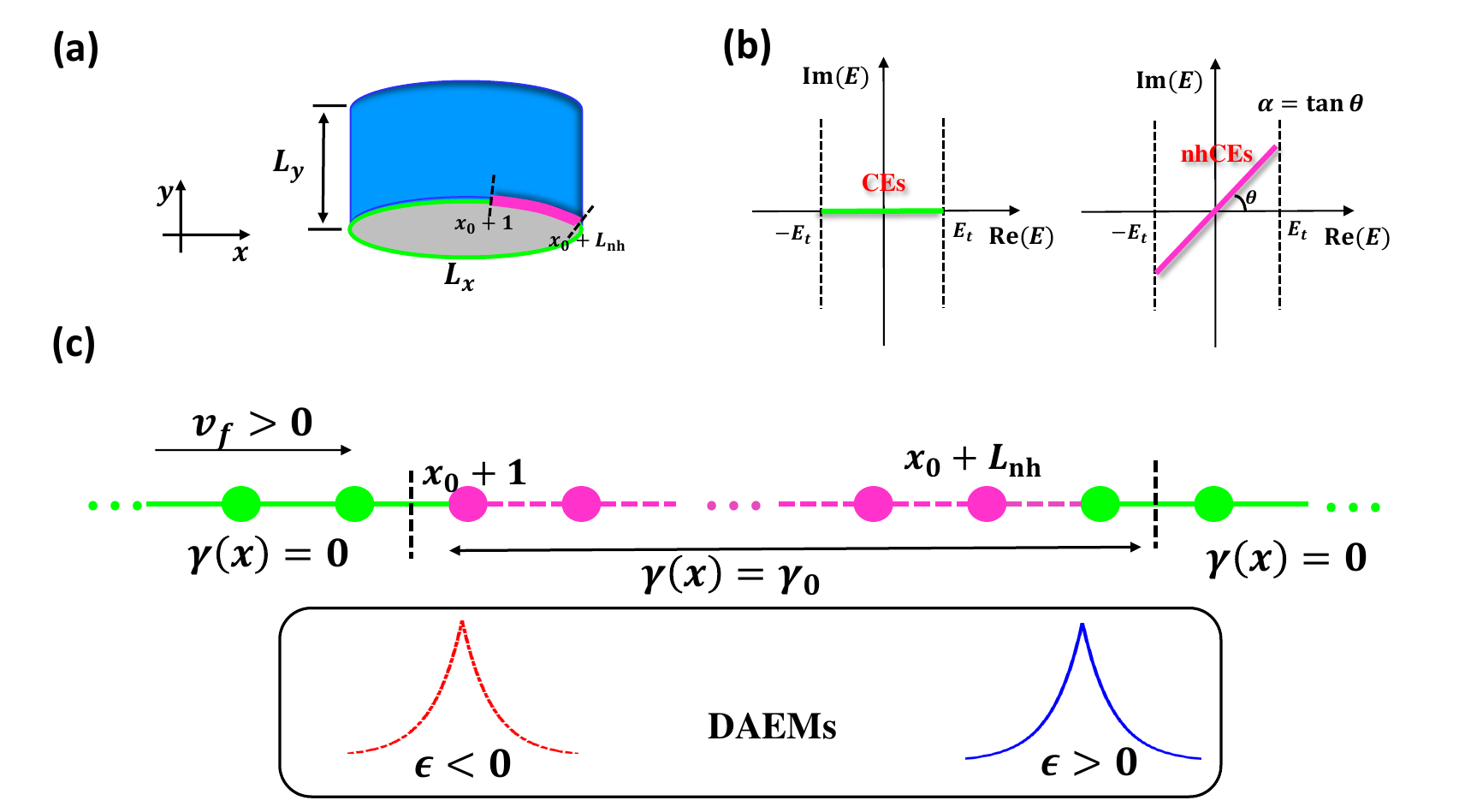}
	\caption{Pictorial illustration of the nhCEs where the dissipation is a Heaviside function at the boundary of a 2D topological materials in the cylinder geometry. (a) The graph of the system in cylinder geometry, where the dissipation is stretches from site $x_0+1$ to site $x_0+L_{\text{nh}}$ of the lower boundary. (b) The shape of complex spectra for the anomalous chiral edge modes (CEs, $|E|<E_{t}$ and $\text{Im}(E)=0$) and the nhCEs ($|\text{Re}(E)|<E_{t}$ and $|\text{Im}(E)|>0$), where $E_{t}=v_f k_t$ corresponds to the bulk gap of system. The complex spectra of nhCEs is a straight line with slope $\alpha=\tan\theta$. (c) The profile of the nhCEs at the lower boundary of topological materials, that the dissipation corresponds to the imaginary Fermi velocity. The edge modes which $\epsilon>0$ are localized at $x_0+L_{\text{nh}}$ and the edge modes which $\epsilon<0$ are localized at $x_0$ if the dissipation. Where $\gamma_0>0$ is assumed.}
	\label{NH_configuration}
\end{figure}

Before we continue our discussion, let's clarify the configuration of topological materials which host the nhCEs. 
The system is in cylinder geometry that periodic in the $x$-direction and open in the $y$-direction, and the dissipation is introduced locally to the lower boundary of system to make its edge modes become dissipative, as present in Fig. \ref{NH_configuration} (a), where the height and the circumference of cylinder are $L_y$ and $L_x$ correspondingly.

For the $k$-dependent dissipation, and the Schr\"{o}dinger equation of the nhCEs reads
\begin{equation}\label{im_velocity}
	\left[ v_{f} + i \gamma(x) \right] \left( -i\frac{d}{dx}\right) \psi(x) = \epsilon\left( 1+i\alpha \right)\psi(x),
\end{equation}
where $\gamma(x)$ and $\alpha$ are real valued, and $i\alpha \epsilon$ corresponds to the imaginary part of the eigenenergy.
The solution of Eq. (\ref{im_velocity}) is
\begin{eqnarray}\label{solu_im_v}
	\psi(x) &=&\frac{1}{\sqrt{R}} \exp\left(\epsilon\int_{0}^{x} dx'\frac{\gamma(x')-\alpha v_{f}}{\gamma^2(x') + v^2_{f}} \right)\nonumber\\ &&\times\exp\left( i\epsilon\int_{0}^{x} dx'\frac{\alpha\gamma(x')+ v_{f}}{\gamma^2(x') + v^2_{f}} \right),
\end{eqnarray}
where $\frac{1}{\sqrt{R}}$ is the normalization factor.
Because the edge of topological materials is a periodic chain in our configuration, then the principle of continuous condition of the wavefunctions implies that $\psi(L_x)=\psi(0)$ should be satisfied.
Then, we have
\begin{equation}
	\epsilon\int_{0}^{L_x}dx'\frac{\left( \gamma(x')-\alpha v_f\right) + i\left( \alpha\gamma(x')+v_f \right)  }{\gamma^2(x')+v^2_f}  = 2i\pi n,
\end{equation}
where $n\in \mathbb{Z}$.
Consequently, the following equations should be satisfied
\begin{eqnarray}
	\alpha = \frac{1}{v_f}\frac{\int_{0}^{L_x}dx'\frac{ \gamma(x') }{\gamma^2(x')+v^2_f}}{\int_{0}^{L_x}dx'\frac{ 1 }{\gamma^2(x')+v^2_f}} = \frac{\tilde{\gamma}}{v_f},\\
	\left( \alpha + \frac{1}{\alpha}\right) \int_{0}^{L_x}dx'\frac{ \gamma(x') }{\gamma^2(x')+v^2_f} = \frac{2\pi n}{\epsilon},
\end{eqnarray}
where $\tilde{\gamma}$ is the average value of $\gamma(x)$ which the weight function is $\frac{1}{\gamma^2(x)+v^2_f}$.
Given that $\alpha$ is a constant when $\gamma(x)$ is fixed, then the real part and the imaginary part of eienenergy is proportional to each other, which means that the complex spectra of these nhCEs is a straight line.

Interestingly, $\psi$ acquires an exponential factor, which implies that the anomalous edge modes might become the localized states.
However, if the sign of factor $\gamma(x)-\alpha v_f$ remains the same along the edge of system, then the edge modes will still be the extended state.
Take $\epsilon>0$ and $\gamma(x)-\alpha v_f>0$ as an example, one can find that $\psi(x)$ is exponentially increase when $x>x_c$, where $1\leq x_c\leq L_x$, which means that there is no site where $\psi(x)$ can localize at.
As the results, the edge modes can only localize at the position where $\gamma(x)-\alpha v_f$ changes sign.
For example, if $\epsilon>0$, $\psi(x)$ can localize at the position $x_c$ that $\gamma(x>x_c)<\alpha v_f$ and $\gamma(x<x_c)>\alpha v_f$, in which $\psi(x)$ is exponentially decay at the both sides of $x_c$; and if $\epsilon<0$, $\psi$ can only localize at the position $x_c$ that $\gamma(x>x_c)>\alpha v_f$ and $\gamma(x<x_c)<\alpha v_f$.

More specifically, let's consider the distribution of dissipation as a Heaviside function at the boundary of system, that
\begin{equation}\label{gamma_distribution}
	\gamma(x)=\left\{\begin{aligned}
		\gamma_0 &\qquad x_0 < x \le x_0+L_{\text{nh}},\\
		0 &\qquad \text{otherwise},
	\end{aligned}
	\right.
\end{equation}
where $L_{\text{nh}}$ is the total links at the boundary of system which have non-reciprocal hopping, or the Fermi velocity of edge modes in these space is a complex number.
For the dissipation in Eq. (\ref{gamma_distribution}), its easy to find that 
\begin{eqnarray}
	\alpha = \frac{\frac{v_f\gamma_0}{v^2_f+\gamma^2_0}L_{\text{nh}}}{  L_x  - \frac{\gamma^2_0}{v^2_f+\gamma^2_0}L_{\text{nh}} }.
\end{eqnarray}
It says that $\alpha$ is fixed when $L_x$, $L_{\text{nh}}$ and $\gamma_0$ are given, which implies that the shape of complex spectra is a straight line.
Owing to continuous condition of the wavefunction, the distribution of $\gamma(x)$ would completely determine the profile of the nhCEs. 
That the modes with the positive sign of $\epsilon$  and those with negative sign of $\epsilon$ are localized at different positions, one is at $x_0$, another one is at $x_0+L_{\text{nh}}$.
Suppose $v_f>0$ and $\gamma_0>0$, then the modes that $\epsilon>0$ are localized at $x_0+L_{\text{nh}}$; while the modes that $\epsilon<0$ are localized at $x_0$, as seen in Fig. \ref{NH_configuration} (b).
While the situation that $v_f<0$ (for the edge modes at another boundary) and $\gamma_0>0$ is the other way around, the modes which $\epsilon>0$ are localized at $x_0$; while the modes which $\epsilon<0$ are localized at $x_0+L_{\text{nh}}$.

Interestingly, the shape of the nhCEs in the space where $\gamma(x)=0$ is exponentially like for the dissipation is a Heaviside function in Eq. (\ref{gamma_distribution}), i.e. $e^{-\kappa x - b}$, and the localization strength $\kappa=\frac{\epsilon\alpha}{v_f}\propto \frac{1}{L_x}\frac{1}{L_x+\beta L_{\text{nh}}}$, where the first factor is because $\epsilon\propto\frac{1}{L_x}$ for the low-energy excitations.
In results, the localization behavior of nhCEs with  complex Fermi velocity is scale-relevant.

In the following sections, we will examine the general law of nhCEs with local complex Fermi velocity in the 2D $p$-wave SC and the Chern insulators, in which the remarkable complex spectra and the wavefunctions' profile of nhCES are investigated.

\section{non-Hermitian chiral majorana edge modes in the 2D $p$-wave SC}
Owing to the superconducting gap, a SC is similar to an insulator, while one thing that distinguishes SC from an insulator is that the former has intrinsic particle-hole symmetry (PHS). The SC is usually described with a Bogoliubov-de Genes Hamiltonian
\begin{equation}
	H_{\text{SC}} = \sum_{\mathbf{k}}\psi^\dagger H_{\text{SC}}(\mathbf{k}) \psi,
\end{equation}
where $\psi$ is a Nambu vector that $\psi = (c_{\mathbf{k}} , c^\dagger_{-\mathbf{k}} )^T$, which is an annihilation operator of both the particles and the holes, and $H_{\text{SC}}(\mathbf{k})$ is a Bloch Hamiltonian.
Because of the PHS, the eigenvector of $H_{\text{SC}}$ is symmetric, that if $\phi_{+}=(u,v)^T$ is the eigenvector with eigenvalue $E$, then $\phi_{-}=(v^*, u^*)^T$ is the eigenvector with eigenvalue $-E$. Meanwhile, the excitations of the system are the Bogoliubov quasiparticles
\begin{equation}
	a^\dagger = u c^\dagger + v c,\qquad a = v^* c^\dagger + u^* c,
\end{equation}
which is composed of both particles and holes.
Similar to topological insulators (TIs), non-trivial bulk topology of $H_{\text{SC}}(\mathbf{k})$ will results in gapless Majorana edge modes, which are the equal superposition of particles and holes ( or $|u|=|v|$ for the Majorana chiral edge modes).
In the 2D $p$-wave SC, the Bloch Hamiltonian is \cite{topo_mat_ref1,topo_mat_ref2}
\begin{eqnarray}\label{Bloch_SC}
	H_{p-\text{SC}} (\mathbf{k}) &=& -2\Delta_0\sin k_y \sigma_x - 2\Delta_0\sin k_x \sigma_y \nonumber\\
	&&+  \left(t\cos k_x + t\cos k_y + \mu -2t\right)\sigma_z,
\end{eqnarray}
where the PHS is: $\sigma_x H^T_{p-\text{SC}} (\mathbf{k}) \sigma_x = -H_{p-\text{SC}} (-\mathbf{k})$.
The topology of 2D $p$-wave SC is captured by the Chern number of the ground state manifold in Eq. (\ref{Bloch_SC}), that the Chern number is non-zero when $0<\mu<4$, and we had set $t=1$ for simplicity.
Amazingly, due to the PHS, the in gap low-energy excitations of a SC are the Majorana fermions, which are their own anti-particles.
These Majorana fermions which have unusual braiding statistical property are key for the topological quantum computation \cite{topo_comp_ref1,topo_comp_ref2}.

When the system is in a cylinder geometry, that the boundary condition along $x$-direction is periodic while the boundary condition along $y$-direction is open.
Then there are Majorana chiral edge modes at the boundaries $y=1$ and $y=L_x$, which is two-component vector
\begin{equation}\label{Majorana_vec}
	\psi_{y=1} \propto [1, i]^T, \qquad \psi_{y=L_y} \propto [1, -i]^T,
\end{equation}
where the internal degrees of freedom in each unit cell are the particles and the holes, that the edge modes are the equal superposition of the particles and the holes, so become the Majorana fermions of the system.
The energy dispersion of the edge modes is
\begin{equation}\label{low_energy_sc}
	E_{p-\text{SC,edge}} = \pm 2\Delta_0 \sin k_x,
\end{equation}
where "$\pm$" specify different boundaries, and the fermi velocity of chiral Majorana edge modes is $v_f = 2\Delta_0$.
Then, we want to figure out how to have the localized Majornan modes by introducing the dissipations, that both the imaginary Fermi velocity and the imaginary potential are discussed.

\subsection{ Two-body gain/loss at the boundary 2D $p$-wave SC}
\begin{figure}[h]
	\includegraphics[width=0.4\textwidth]{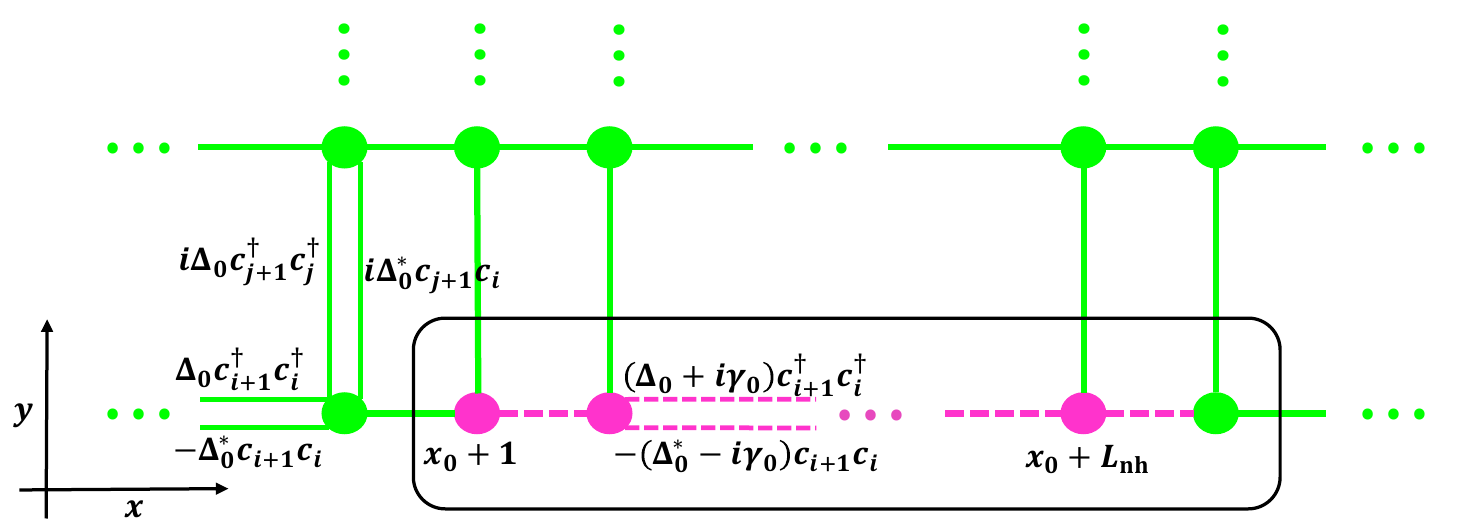}
	\caption{Pictorial illustration of the 2D $p$-wave superconductor that experiencing two-body gain at its lower boundary, where two-body gain is for the links that are marked with magenta dashed line which stretches from site $x_0+1$ to site $x_0+L_{\text{nh}}$.}
	\label{SC_pic}
\end{figure}

In the 2D $p$-wave SC, the momentum part in the corresponding Dirac Hamiltonian is $\Delta_0\sin k_x \sigma_y$.
The Fermi velocity of the Majorana edge mode is the pairing term along $x$-direction, and the complex Fermi velocity corresponds to the pairing amplitude being a complex number $\Delta_0+i\gamma_0$.
This can be realized with the two-body gain (loss) along the boundary, which is usually caused by the inelastic collisions \cite{two_body_loss1,two_body_loss2}.
Then, suppose that the dissipation is a Heaviside function in Eq. (\ref{gamma_distribution}), that the links among the lattices that range from site $x_0+1$ to site $x_0+L_{\text{nh}}$ experiencing the two-body gain, as presented in Fig. \ref{SC_pic}.

\begin{figure}[h]
	\includegraphics[width=0.5\textwidth]{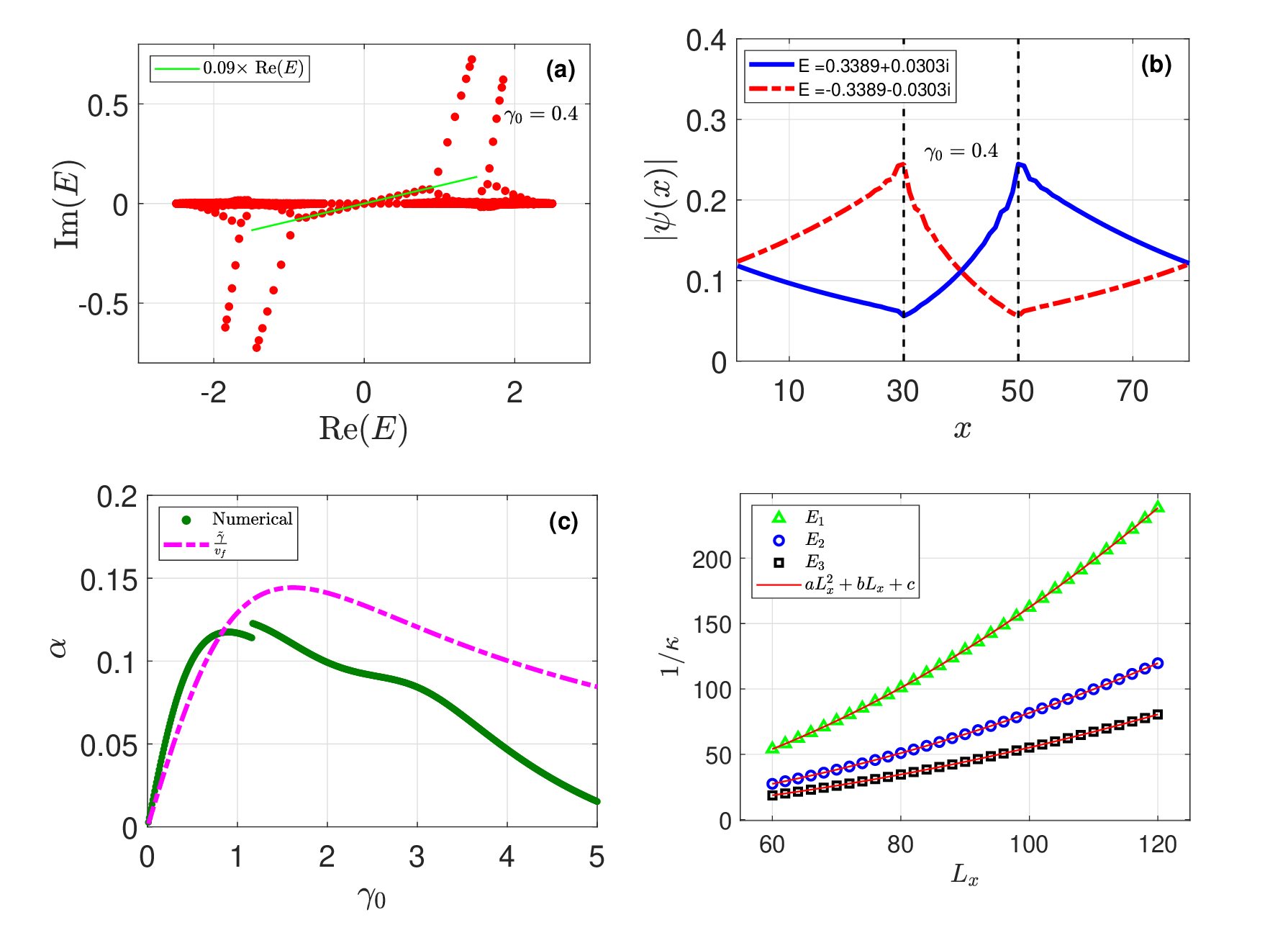}
	\caption{(a) The complex spectra of 2D $p$-wave SC which there are $L_{\text{nh}}=20$ links that range from site $31$ to site $50$ at the boundary experience two-body gain. (b) The corresponding wavefunctions of the NH Majorana edge modes, that $|\psi(x)| = |u(x)|+|v(x)|$. (c) The slope of complex spectra line $\alpha$ versus $\gamma_0$, where the magenta dash-dotted line is the prediction value of $\alpha$. (d) The inverse localization strength $1/\kappa$ versus $L_x$ for the nhCEs with imaginary Fermi velocity in the 2D $p$-wave SC. Where the localization strength $\kappa$ is obtained by fitting the wavefunctions of edge modes with $e^{-\kappa x - \beta}$ in the space where $\gamma(x)=0$. In which $E_1$, $E_2$ and $E_3$ are the modes which $\text{Re}(E)>0$ and $|\text{Im}(E)|>0$, and they are arranged by their real part in the ascend order for different values of $L_x$, i.e. $\text{Re}(E_1)<\text{Re}(E_2)<\text{Re}(E_3)$. Where $L_x=80$, $L_y=20$, $\Delta_0=0.7$ and $\mu=1.5$.}
	\label{SC_im_v}
\end{figure}

The complex spectra of 2D $p$-wave SC which its boundary experiencing two-body gain is presented in Fig. \ref{SC_im_v} (a), which the complex spectra of nhCEs is a straight line.
Moreover, if the dissipation corresponds to the imaginary Fermi velocity of anomalous edge modes, then the edge modes are localized at $x_0$ and $x_0+L_{\text{nh}}$, see Fig. \ref{SC_im_v} (b), that the modes with $\text{Re}(E)>0$ are localized at $x_0+L_{\text{nh}}$, while the modes with $\text{Re}(E)<0$ are localized at $x_0$.

Moreover, in Fig. \ref{SC_im_v} (c), the numerical value of $\alpha$ is obtained by fitting the complex spectra of nhCEs in the region $|\text{Re}(E)|<0.5$ and $|\text{Im}(E)|>0$, this is because that the bulk gap of 2D $p$-wave SC in Eq. (\ref{Bloch_SC}) is $\Delta_g = 0.5$ when $\mu=1.5$.
We find that $\alpha$ is well described with $\frac{\tilde{\gamma}}{v_f}$, however it is much smaller than $\frac{\tilde{\gamma}}{v_f}$ when $\gamma_0$ is sufficient high, see the situation that $\gamma_0>1$ in Fig. \ref{SC_im_v} (c).
This is attribute to the bound state, that the effective dissipation $\gamma_e$ that the edge modes felt is smaller than $\gamma_0$ if there is bound state, such that $\alpha$ is smaller than $\frac{\tilde{\gamma}}{v_f}$.

Furthermore, the scaling behavior of the nhCEs with complex Fermi velocity is checked in Fig. \ref{SC_im_v} (d), that the inverse localization strength $\frac{1}{\kappa}$ in the space where $\gamma(x)=0$ for different values of $L_x$ is well described by $aL_x^2 + bL_x +c$.
The scaling behavior of the wavefunctions is originates from two parts, one is that $\alpha\propto\left( L_x+\beta L_{\text{nh}}\right) ^{-1}$, another one is that $\epsilon\propto L_x^{-1}$, such that $\frac{1}{\kappa}=\frac{v_f}{\epsilon\alpha}=aL_x^2 + bL_x +c$.

\subsection{gain/loss of particles at the boundary of 2D $p$-wave SC}
\begin{figure}[h]
	\includegraphics[width=0.5\textwidth]{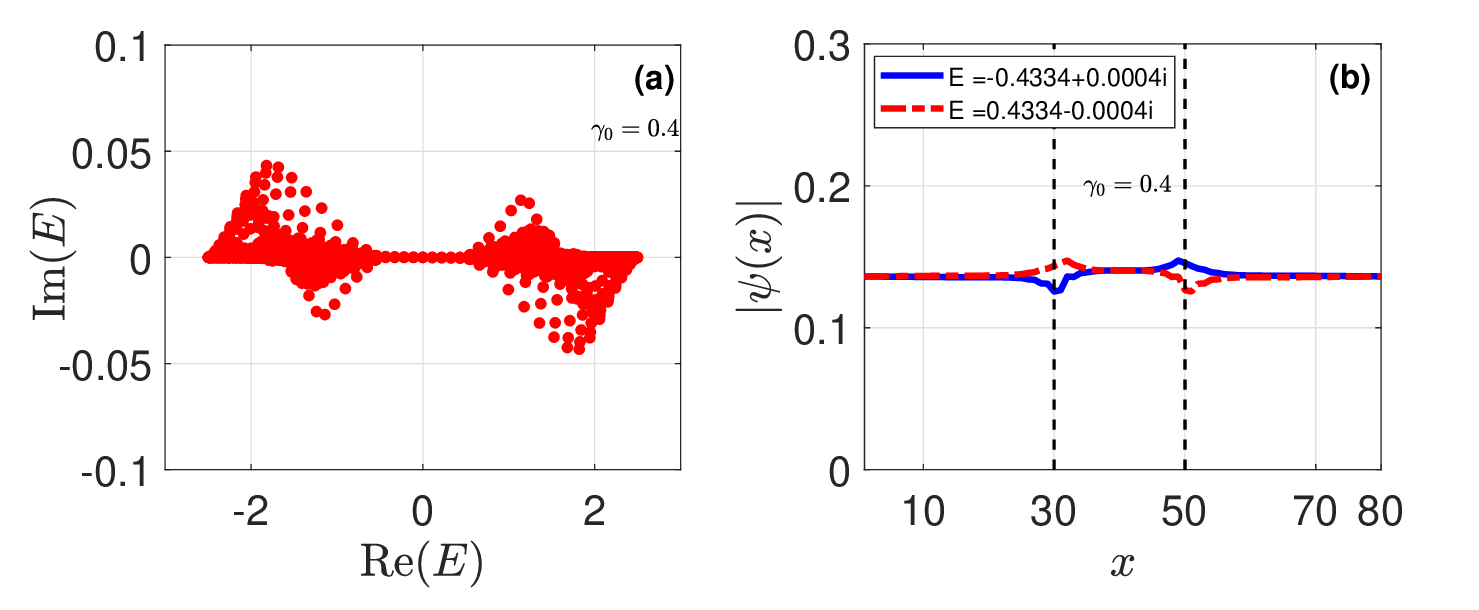}
	\caption{(a) The complex spectra of 2D $p$-wave SC that there is gain of particle among the $L_{\text{nh}}=20$ lattices that range from site $31$ to site $50$ at the boundary of system. (b) The corresponding wavefunctions of the non-hermitian Majorana edge modes, that $|\psi(x)| = |u(x)|+|v(x)|$. Where $L_x=80$, $L_y=20$, $\Delta_0=0.7$ and $\mu=1.5$. }
	\label{SC_im_p}
\end{figure}

It is proposed in Ref. \cite{chiral_localization} that the nhCEs with local imaginary potential become the localized sates, which the dissipation corresponds to the local gain/loss of particles.
However, local gain/loss of particles is not as usual as we expected in SC, the localized Majornana edge modes was not observed by introducing an on-site imaginary potential at the boundary of a 2D $p$-wave SC. Instead the nhCEs remain extended, as presented in Fig. \ref{SC_im_p} (b).
And the in-gap Majornana edge modes acquire a tiny imaginary part, see Fig. \ref{SC_im_p} (a), this is because that the on-site imaginary potential can still perturb its eigenenergy.

The reason why nhCEs remain extended for such occasion is because that the on-site imaginary potential keeps the PHS of SC \cite{NH_majorana}.
Because the Majornana chiral edge modes are the equal superposition of particles and holes, and the gain of particles is also the loss of holes, so the net contribution of imaginary potential to the Majornana chiral edge modes is zero as long as the PHS is maintained.
In other words, the gain/loss of particles is not the imaginary potential to anomalous edge modes of the topological SC, therefore the nhCEs remain extended.
If the PHS is broken, that the dissipation corresponds to the gain/loss of particles only or the gain/loss of holes only, then the chiral localization of the dissipative Majornana edge modes would be possible.
However, it is impossible to break the PHS of SC because it is an intrinsic symmetry, so we conclude that the nhCEs remain extended in 2D $p$-wave SC when the dissipation is the local gain/loss of particles.

\section{ Non-Hermitian chiral edge modes in the Chern insulators}
The anomalous chiral edge modes in a Chern insulator are act as dissipationless channel that transports current along the boundary of system, and it is restricted to moving in a single direction.
These modes are protected by the bulk topology of system and are robust against the disorder and the impurity.
However, it is not known what will happen to these chiral edge modes when the dissipation corresponds to the imaginary Fermi velocity is introduced.

Take the 2D QWZ model as an example, in which the Bloch Hamiltonian is \cite{topo_mat_ref1,topo_mat_ref2}
\begin{eqnarray}\label{qwz_bloch}
	H_{\text{qwz}}(\mathbf{k}) &=& t_x\sin k_x \,\sigma_x + t_y\sin k_y \,\sigma_y \nonumber\\
	&&+ \left( u + t_x\cos k_x + t_y\cos k_y\right)\sigma_z,
\end{eqnarray}
where the Pauli matrices $\sigma_{x,y,z}$ represent the internal degrees of freedom in each lattice site, which denote as $A$ and $B$, $t_x$ ($t_y$) is the hopping amplitude along the $x$ ($y$)-direction, and $u$ is the on-site potential which determines the band-gap $\Delta_g$ of system. 
For QWZ model, edge long-wave-length excitations exist when the Chern number of its ground state manifold is non-zero. This is satisfied when $-2<u<2$, where we had set $t_x=t_y=1$ for simplicity.
When the QWZ model is in a cylinder geometry as illustrated in Fig. \ref{NH_configuration} (a), there are two chiral edge modes at the boundaries $y=1$ and $y=L_y$ that propagate in opposite directions along the boundaries (the $x$-direction).
The edge modes are the two-component wavefunctions, referred to as
\begin{equation}
	\psi_{y=1} \propto [ 1, 1]^T, \qquad \psi_{y=L_y} \propto [ 1, -1]^T,
\end{equation}
where the basis is the internal degrees of freedom $A$ and $B$.
The energy dispersion of the edge modes is
\begin{equation}\label{low_energy_qwz}
	E_{\text{qwz,edge}}(k_x) = \pm t_x \sin k_x,
\end{equation}
where the sign "$\pm$" depends on different boundaries, that the Fermi velocity of chiral edge modes is $t_x$, i.e. $v_f=1$.

\begin{figure}[h]
	\includegraphics[width=0.4\textwidth]{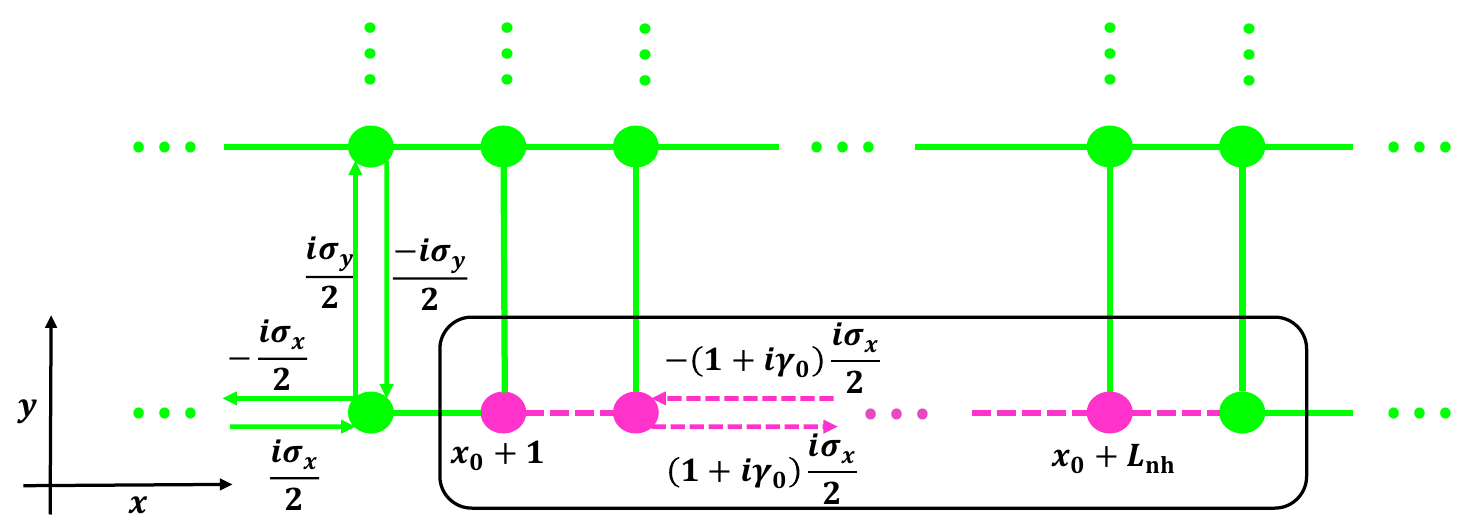}
	\caption{Pictorial illustration of 2D QWZ model which has the non-reciprocal hopping at its lower boundary, where the non-reciprocal hopping is for the links that are marked in magenta dashed line which stretches from site $x_0+1$ to site $x_0+L_{\text{nh}}$.}
	\label{QWZ_pic}
\end{figure}

Now, we introduce the dissipation to the boundary of 2D QWZ model to make the Fermi velocity of edge modes becomes a complex number, that there are $L_{\text{nh}}$ ($L_{\text{nh}}<L_x$) links that stretches from site $x_0+1$ to site $x_0+L_{\text{nh}}$ which the hopping amplitude is $1+i\gamma_0$.
Given that the momentum part is $t_x\sin k_x \sigma_x$ in the context of Dirac Hamiltonian in Eq. \ref{qwz_bloch}, then to meet with our expectation, it is necessary that the hopping between $(x,A)$ and $(x+1,B)$ and the hopping between $(x,B)$ to $(x+1,A)$ is a complex number $1+i\gamma_0$, as presented in Fig. \ref{QWZ_pic}.
\begin{figure}[h]
	\includegraphics[width=0.5\textwidth]{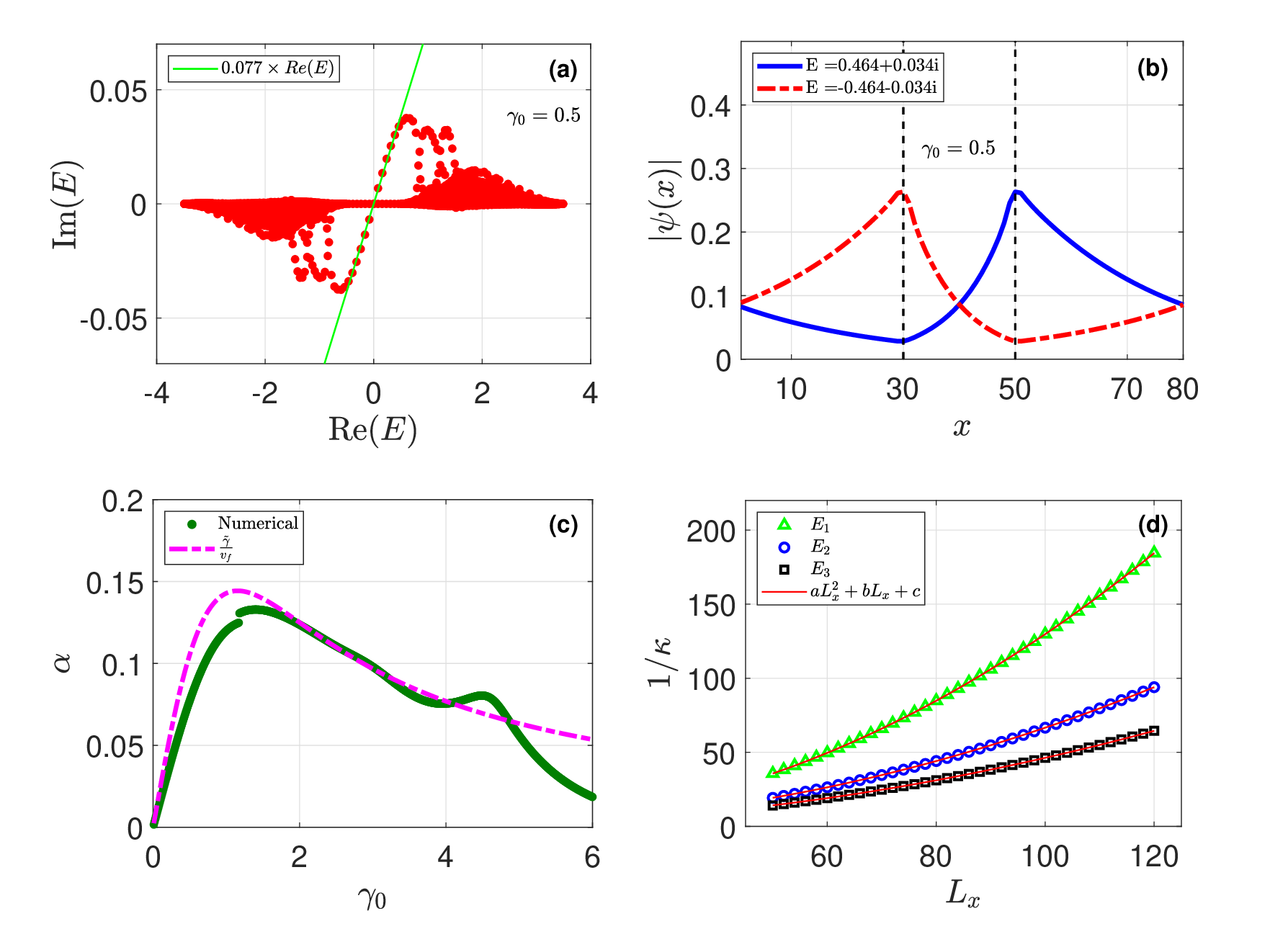}
	\caption{(a) The complex spectra of 2D QWZ model that there are $L_{\text{nh}}=20$ links that range from site $31$ to site $50$ at the boundary which have non-reciprocal hopping. (b) The corresponding wavefunctions of the nhCEs, where $|\psi(x)|=|\psi_A(x)|+|\psi_B(x)|$. (c) The slope of complex spectra line $\alpha$ versus $\gamma_0$, where the magenta dash-dotted line is the prediction value of $\alpha$. (d) The inverse localization strength $1/\kappa$ versus $L_x$ for the nhCEs with imaginary Fermi velocity in the QWZ model. Where the localization strength $\kappa$ is obtained by fitting the wavefunction of edge modes with $e^{-\kappa x - \beta}$ in the space where $\gamma(x)=0$. And $E_1$, $E_2$ and $E_3$ are the modes which $\text{Re}(E)>0$ and $|\text{Im}(E)|>0$, and they are arranged by their real part in the ascend order for different values of $L_x$, i.e. $\text{Re}(E_1)<\text{Re}(E_2)<\text{Re}(E_3)$. Where $L_{x}=80$, $L_{y}=20$, and $u=1.5$.}
	\label{qwz_im_v}
\end{figure}

The complex spectra of 2D QWZ model that has non-reciprocal hopping at its lower boundary $y=1$ is presented in Fig. \ref{qwz_im_v} (a), which the complex spectra of nhCEs is a straight line, i.e. $\text{Im}(E) = \alpha\times\text{Re}(E)$ is satisfied for the nhCEs, this is because that $\alpha=\frac{\tilde{\gamma}}{v_f}$ is fixed when $\gamma_0$, $L_x$ and $L_{\text{nh}}$ are given.
And the profile of the wavefunctions for the nhCEs at the boundary $y=1$ are presented in Fig. \ref{qwz_im_v} (b), that the modes which $\text{Re}(E)<0$ are localized at site $x_0$; while the modes which $\text{Re}(E)>0$ are localized at site $x_0+L_{\text{nh}}$.

Moreover, in Fig. \ref{qwz_im_v} (c), the numerical value of $\alpha$ is obtained by fitting the complex spectra of nhCEs in the region $|\text{Re}(E)|<0.5$ and $|\text{Im}(E)|>0$, this is because that the bulk gap of QWZ model in Eq. (\ref{qwz_bloch}) is $\Delta_g = 0.5$ when $u=1.5$.
We find that $\alpha$ is well described with $\frac{\tilde{\gamma}}{v_f}$.
And the scaling behavior of the nhCEs with complex Fermi velocity in the QWZ model is checked in Fig. \ref{qwz_im_v} (d), that the inverse localization strength $\frac{1}{\kappa}$ in the space where $\gamma(x)=0$ for different values of $L_x$ is well described by $aL_x^2 + bL_x +c$.

\section{conclusion and discussion}
In this paper, we investigated the properties of NH anomalous chiral edge modes that has local complex Fermi velocity.
By solving the NH Schr\"{o}dinger equation, we find that these dissipative modes that reside on the boundary of topological materials become the localized states.
We find that the complex spectra of these nhCEs is a straight line.
Surprisingly, the chirality of edge modes will determine the position where the nhCEs are localized, that the positive energy modes and the negative energy modes are localized at different positions.
Different from the scale-free localization that induced by the local non-reciprocal hopping in the 1-dimension chain \cite{local_nh_PT}, the localization of nhCEs with complex Fermi velocity is scale-relevant, which is the key property of low-energy physics.
Furthermore, the general properties of these nhCEs have been tested in the 2D topological materials, the 2D $p$-wave SC and the QWZ model, which is consistent with our theoretical prediction.

Significantly, previous studies on the nhCEs with local imaginary potential \cite{chiral_localization} cannot be generalized to the 2D $p$-wave SC, as there are no localized edge modes when the dissipation is local gain/loss of particles.
This is because that the intrinsic PHS is maintained if there is on-site imaginary potential for topological SC \cite{NH_majorana}.
The Majorana edge modes in the 2D $p$-wave SC are equal superposition of particles and holes, while the gain of particles is also loss of holes, so the net contribution of imaginary potential to the Majorana chiral edge modes in the 2D $p$-wave SC  is zero.
As a results, it is impossible to introduce an imaginary potential to the chiral Majorana edge modes, then we can't localize Majorana edge modes by introducing a local gain/loss of particles.
This means that the only way to have localized Majorana edge modes in the dissipative context is for the dissipation to play the role of an imaginary Fermi velocity for the chiral Majorana edge modes.

In the NH system, The localization of the wavefunction is related to a unique phenomena called the NH skin effect.
The physical implications of the NH skin effect have been extensively studied, such as chiral damping in open quantum system \cite{chiral_damping}.
It is expected that the localization of nhCEs would have its counterpart in the corresponding dynamic process, which would alter the transport property of the anomalous chiral edge modes.
This is worth of researching and is a potential future direction.

\section{acknowledgment}
This work is supported by NSFC Grant No. 11974053 and No. 12174030.

\end{document}